\documentclass[reprint,apshs,prb]{revtex4-1}
\usepackage{amsmath,amssymb}
\usepackage{bm}
\usepackage{graphicx}
\usepackage{subfigure}
\usepackage{slashbox}
\usepackage{tabularx}

\newcommand{\mA}{{\mathsf A}}
\newcommand{\mB}{{\mathsf B}}

\newcommand{\neiN}{{\mathfrak n}}

\newcommand{\cut}{{\rm cut}}

\newcommand{\qm}{{\rm qm}}

\newcommand{\btheta}{{\bm\theta}}

\newcommand{\transpose}{{\!\top}}

\begin{document}

\title{Accelerating high-throughput searches for new alloys with active learning of interatomic potentials}
\author{Konstantin Gubaev, Evgeny V. Podryabinkin, \\ Gus L. W. Hart$^*$, Alexander V. Shapeev$^*$}

\begin{abstract}
We propose an approach to materials prediction that uses a machine-learning interatomic potential to approximate quantum-mechanical energies and an active learning algorithm for the automatic selection of an optimal training dataset.
Our approach significantly reduces the amount of DFT calculations needed, resorting to DFT only to produce the training data, while structural optimization is performed using the interatomic potentials.
Our approach is not limited to one (or a small number of) lattice
types (as is the case for cluster expansion, for example) and can predict structures with lattice types not present in the training dataset.
We demonstrate the effectiveness of our algorithm by predicting the convex hull for the following three systems: Cu-Pd, Co-Nb-V, and Al-Ni-Ti.
Our method is
three to four orders of magnitude faster than conventional high-throughput DFT calculations and
  explores a wider range of materials space.
In all three systems, we found
unreported stable structures compared to the AFLOW database.
Because our method is much cheaper and explores much more
of materials space than high-throughput methods or cluster
expansion, and because our interatomic potentials have a systematically improvable accuracy compared to empirical potentials such as EAM, it
will have a significant impact in the discovery of new alloy phases,
particularly those with three or more components.
\end{abstract}

\maketitle

\section{Introduction}
Advances in computer power, improvements in first-principles methods,
and the generation of large materials databases like AFLOWLIB
  \cite{curtarolo2012aflowlib}, OQMD \cite{saal2013materials}, CMR
  \cite{landis2012computational}, NOMAD \cite{nomad}, and Materials
Project\cite{Jain2013} have enabled modern data analysis tools to be
applied in the field of materials discovery
\cite{kalidindi2015materials,agrawal2016perspective,rajan2015materials}.
There have been growing efforts in computational search for materials
with superior properties, including metallic alloys
\cite{nyshadham2017computational,curtarolo2012aflow,hart2013comprehensive},
semiconductor materials \cite{hinuma2016discovery}, and magnetic
materials \cite{sanvito2017accelerated}.  In this work we consider the
problem of predicting stable phases in multicomponent alloys.
A typical prediction algorithm consists of
sampling structures across the configurational space and evaluating
their energies.  The sampling is done by searching through structures
that are either selected from some carefully assembled pool of
possible structures, often called crystal prototypes
\cite{mehl2017aflow}, or are generated by some sampling algorithm,
see, e.g., Refs.~\onlinecite{hart2012generating,lyakhov2013new}

The evaluation of the energy of the structures in the pool is often
done with density functional theory (DFT).  Even despite its favorable
accuracy/efficiency trade-off as compared to other quantum-mechanical
algorithms, the DFT calculations remain the bottleneck in materials prediction workflows, making an exhaustive search
impractical. Machine learning (ML) for materials prediction has
the potential to dramatically reduce the number of quantum-mechanical
calculations performed and thus reduce the computational expense of
predicting new materials via computation.  The reduction of the
computational time is achieved by constructing a surrogate model that
``interpolates'' the quantum-mechanical training data and makes
subsequent energy evaluations much faster (by orders of
magnitude). This is similar in spirit to the cluster expansion method
which has been broadly used in different materials discovery
applications
\cite{wu2016cluster,jiang2016efficient,troparevsky2015beyond,hinuma2016discovery}. Cluster
expansion is quite successful when the stable structures are
derivatives of a particular structure (fcc, bcc, etc.) but is not
useful when this is not the case. Its accuracy also converges slowly
when atomic size mismatch is not negligible
\cite{PhysRevB.96.014107}. Additionally, more classical
machine-learning algorithms such as decision trees
\cite{meredig2014combinatorial}, support vector machines
\cite{ubaru2017formation}, and other ML algorithms
\cite{Shapeev2016hea,ward2016general} have been tried. Surrogate
models such as the cluster expansion and standard machine learning
approaches do not have the broad applicability and exceptional
accuracy of the MTP-based active learning approach we demonstrate
here.

The two important features of our approach are a completely general
form for the interatomic potentials and an active learning algorithm
for generating and refining the training set.
In this work we extend the approach \cite{podryabinkin2018accelerating} for predicting the structure of a single-component material.
In our approach, a ML model reproduces DFT for off-equilibrium structures that are not restricted to any lattice.  Furthermore, the model learns the DFT interaction actively (on-the-fly) while equilibrating the candidate structures, completely automating the construction of the training set.
Thus, structural
optimization of the training structures can be performed via the
interatomic potentials, rather than via DFT, further accelerating the
construction of the training set.

Our method is based on moment tensor potentials (MTPs \cite{shapeev2016moment}) and the active learning algorithm \cite{podryabinkin2017active}.
Namely, we solve the following problem: given a set of elements, find the most stable structures (in the sense of lying on the convex hull of formation enthalpies) consisting of these elements, characterized by their composition, unit cell geometry and atomic positions within the unit cell. In this work we extend the interatomic potential \cite{shapeev2016moment} and active learning algorithm \cite{podryabinkin2017active} to handle atomistic configurations with multiple types of atoms, similarly to the approach used in cheminformatics \cite{gubaev2018machine}.The differences between the algorithms from Ref.~\onlinecite{gubaev2018machine} and this work include that (1) we need derivatives of the energy, whereas in Ref.~\onlinecite{gubaev2018machine} we needed only the energy (or other predicted properties); and (2) that in Ref.~\onlinecite{gubaev2018machine} we were concerned with a selection from a finite set of predefined structures, whereas in this work we need to solve the problem predicting the energy with a fitted potential and assembling the training set used for the fitting at the same time (in other words, exploring the potential energy landscape and constructing the training set at the same time).

	The idea of applying neural networks, as a broad class of machine-learning algorithms, to constructing interatomic potentials was pioneered in Ref.~\onlinecite{Behler}.
	Application of Gaussian process regression, another class of machine-learning algorithms, was then proposed in Ref. \cite{Bartok}.
	The promising results obtained in these works have motivated many research groups to pursue this research direction \cite{ArtrithKolpak2015NNP,Behler2011NNP,0953-8984-26-18-183001,Boes2016NNP-ReaxFF-comparison,Dolgirev2016,Gastegger2015high,Manzhos2015neural,NatarajanMorawietzBehler2015water,Lubbers2018,Smith2017,Kolb2017,GAP2014,DeringerCsanyi2016carbon,Deringer2018Boron,Grisafi2018tensorial,Thompson2015316,BotuRamprasad2015-MLIP,LiKermodeDevita2015lotf,Kruglov2017energy-free,sGDML,shapeev2016moment,schutt2017schnet}.
	However, the application of such algorithms to the problem of materials prediction has proven difficult since such a methodology requires one to collect all the representative structures in the training set which is as hard as predicting materials structure itself.
	In our view, it is the active learning \cite{podryabinkin2017active,podryabinkin2018accelerating,0953-8984-26-18-183001,Botu2016machine,smith2018less-is-more} that paves the way for machine-learning interatomic potentials to accelerating computational materials discovery.

This paper is organized as follows: in Section \ref{sec:meth} we introduce the algorithms we use, including the moment tensor potentials (Section \ref{sec:meth:mlip}), active learning (Section \ref{sec:maxV}), and the ``relaxation while learning on-the-fly'' algorithm (Section \ref{sec:lotf}).
In Section \ref{sec:results-and-discussion} we test the proposed algorithm on predicting the stable structures of the Cu-Pd, Co-Nb-V, and Al-Ni-Ti systems and discuss the performance of our algorithm. In particular, we compare our results to those obtained by high-throughput DFT calculations as reported in the AFLOW database \cite{curtarolo2012aflow,curtarolo2012aflowlib}.
In all three systems we have discovered new structures below the reported convex hull of ground-state structures.
Finally, in Section \ref{sec:conclusions} we make concluding remarks.

\section{Methodology} \label{sec:meth}

\subsection {Machine-learning potentials} \label{sec:meth:mlip}

We use the moment tensor potentials (MTPs) for approximating a quantum-mechanical energy.
The potential is parametrized by a set of parameters $\btheta$ that are found from minimizing the loss functional expressing that the predicted energy $E$ is close to the reference quantum-mechanical energy $E^\qm$:
\begin{equation}\label{eq:loss}
L(\btheta)=\sum_j \Big(E\big(\btheta,x^{(j)}\big) - E^\qm\big(x^{(j)}\big)\Big)^2 \longrightarrow \min,
\end{equation}
where $x^{(j)}$ are the configurations in the training set and $E^\qm\big(x^{(j)}\big)$ are their reference energies.

\begin{figure}[htbp] 
	\includegraphics[width=\linewidth]{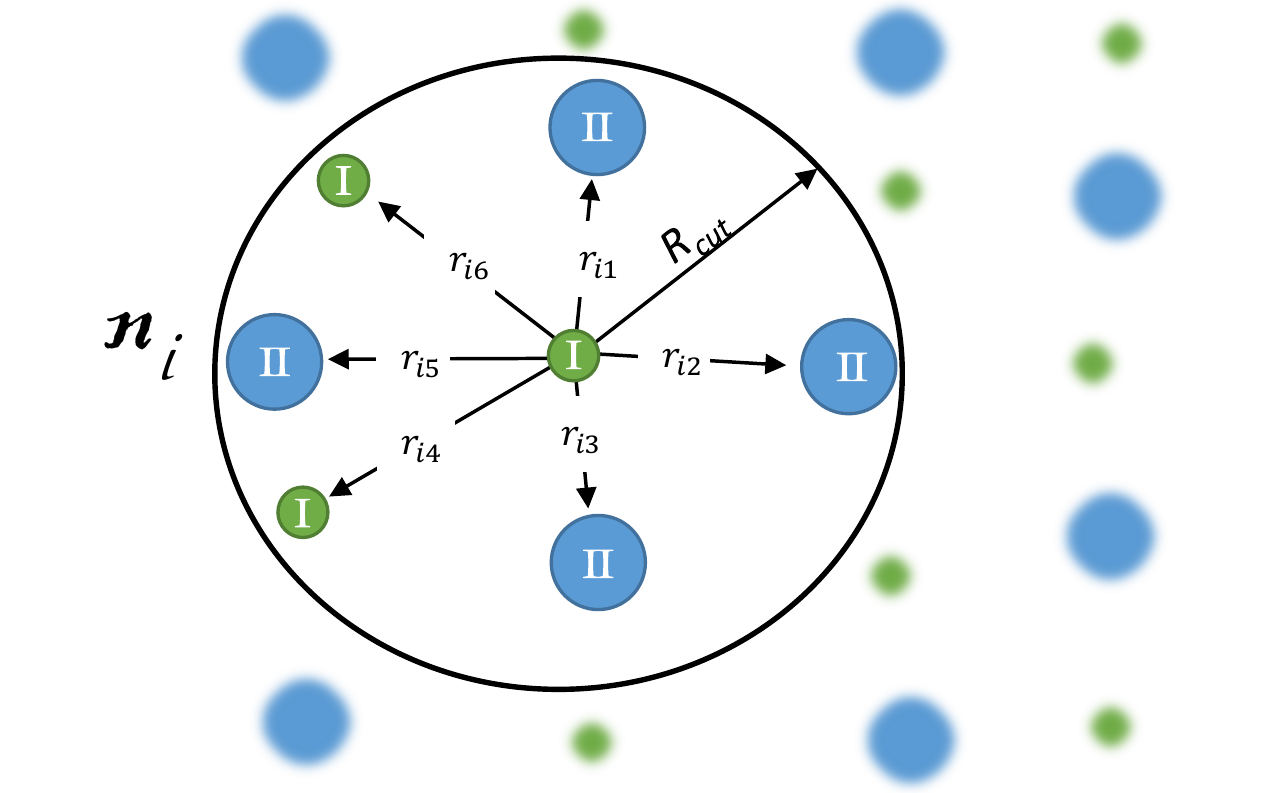}
	\caption {Partitioning scheme: energy $E$ is composed from contributions $V_i$ of individual neighborhoods $\neiN_i$. The neighborhood $\neiN_i$ of the $i$th atom is described by the relative position of neighboring atoms, $\bm r_{ij}$, and the types of atoms $z_j$ (I or II in this illustration).}
	\label{fig:local}
\end{figure}

Our model is local, which we enforce by partitioning the energy, $E$,
into the contributions, $V$, of individual atomic neighborhoods.  To
define a neighborhood of the $i$th atom, we let $\bm r_{ij}$ be the
position of $j$th atom relative to the $i$th atom (thus, $\bm r_{ij}$
is a vectorial quantity)
and $z_j$ be the type of the $j$th atom.  Then $\neiN_i$ is the collection of $\bm r_{ij}$ and $z_j$, and $E(x) = \sum_i V(\neiN_i)$.
The locality of the model is expressed by the requirement that $V$ does not depend on atoms that are farther from $i$ than some cutoff distance $R_{\cut}$, which is usually around $5$ \AA.  An illustration of an atomic neighborhood is sketched in Figure \ref{fig:local}.
%
Mathematically, each atom in the neighborhood introduces four degrees of freedom, on which $\neiN_i$ depends: these are three coordinates in Euclidean space, and a discrete variable representing the chemical type. Typically, neighborhoods include a few dozen atoms, which means that the function $V(\neiN_i)$ depends on the order of hundred scalar variables.
To somewhat reduce the dimensionality, we embed all the physical symmetries into $V(\neiN)$ so they will not have to be learned by the model. 
These symmetries arise from the isotropy and translational symmetry of the physical space, and from the fact that the interaction does not depend on the ordering of atoms. 

As in the work \cite{shapeev2016moment} devoted to the single-component moment tensor potentials, $V(\neiN)$ is linearly expanded through a set of \textit{basis functions} $B_\alpha$:
\begin{equation} \label{eq:lin_expansion}
V(\neiN) = \sum_\alpha \xi_\alpha B_\alpha(\neiN).
\end{equation}
The basis functions, in turn, depend on the set of \textit{moment tensor descriptors}
%
%
\begin{equation}\label{eq:moments}
M_{\mu,\nu}(\neiN_i)=\sum_{j} f_{\mu}(|\bm r_{ij}|,z_i,z_j)\underbrace {\bm r_{ij}\otimes...\otimes \bm r_{ij}}_\text{$\nu$ times},
\end{equation}
where the index $j$ enumerates all the atoms in the neighborhood $\neiN_i$.
The functions $f_{\mu}(|\bm r_{ij}|,z_i,z_j)$ depend only on the interatomic distances and atomic types, therefore we call them \textit{radial functions}. 
The terms $\bm r_{ij}\otimes...\otimes \bm r_{ij}$ contain the angular
information about the neighborhood $\neiN_i$ and are tensors of rank
$\nu$.
We next explain how to construct the basis functions from the moment tensor descriptors, following which we present a simple illustration of the structure of the descriptors and basis functions.


The functions $B_\alpha(\neiN_i)$ enumerate all possible contractions of any number of $M_{\mu,\nu}(\neiN_i)$ yielding a scalar.
Note that $M_{\mu,\nu}(\neiN_i)$ are invariant, by construction, with respect to translations of the system and permutations of equivalent atoms. Their scalar contractions are invariant with respect to rotations of the neighborhood. Thus the resulting function $V(\neiN)$ also has these symmetries.
Although all the descriptors $M_{\mu,\nu}(\neiN_i)$ are composed of two-body terms depending only on $\bm r_{ij}$, their contractions $B(\neiN_i)$ can depend on many-body terms of higher order.

For the purpose of illustration, assume, for the moment, that
the vectors $\bm r_{ij}$ are two-dimensional and that we can express them in polar coordinates $(\rho,\theta)$ centered at the $i$th atom.
Let us look closer at the term $\bm r_{ij}\otimes...\otimes \bm r_{ij} =: r^{\otimes \nu}$.
$\bm r_{ij}^{\otimes 0}$ is a scalar with no angular information, while $\bm r_{ij}^{\otimes 1} = \bm r_{ij} = |\bm r_{ij}| (\cos\theta_{ij},\sin\theta_{ij})$ does contain angular information.
A vectorial contraction is simply a
dot product: $\bm r_{ij}\cdot \bm r_{ik} = |\bm r_{ij}|\,|\bm r_{ik}|
\cos(\theta_{ij}-\theta_{ik})$---in this way we introduce angular terms into the potential.
An arbitrary function of angle can be expanded into a sum of powers of cosine.
Such higher-order terms are contributed to the potential by higher-rank tensors, e.g.,
\[
\bm r_{ij}^{\otimes 2} = \bm r_{ij} \bm r_{ij}^{\transpose}
= |\bm r_{ij}|^2 \begin{pmatrix}
\cos^2\theta_{ij} & \sin\theta_{ij} \cos\theta_{ij} \\
\sin\theta_{ij} \cos\theta_{ij} & \sin^2\theta_{ij}
\end{pmatrix}.
\]
The contractions of two matrices are given by the Frobenius product
\[
\bm r_{ij}^{\otimes 2} \!:\! \bm r_{ik}^{\otimes 2}
= |\bm r_{ij}|^2 |\bm r_{ik}|^2 \cos^2(\theta_{ij}-\theta_{ik}).
\]
A more complicated expression can be constructed with a matrix and two vectors:
\[
(\bm r_{ij}^{\otimes 2}  \bm r_{ik})\cdot \bm r_{i\ell}
= |\bm r_{ij}|^2 |\bm r_{ik}| |\bm r_{i\ell}|
\cos(\theta_{ij}-\theta_{ik}) \cos(\theta_{ij}-\theta_{i\ell}).
\]
Terms of this form are rotationally invariant.
Permutation invariance is achieved by summing those terms over all atoms in the neighborhood weighted by the radial functions.

As an illustration, assume that we have two radial functions,
\[
f_\mu(\rho,z_i,z_j) = \exp\Big(-\frac{|\rho-R_\mu|^2}{2\sigma^2}\Big),
\]
$\mu=1,2$, where $\rho$ has the meaning of distance to the central, $i$th atom.
In the sum \eqref{eq:moments} they ``extract'' two shells of atoms,
around the distances $R_1$ and $R_2$ from the $i$th atom, smeared over
the width of $\sigma$.
We did not, but could assume the dependence of these functions on the types of atoms $z_i$ and $z_j$---this would discriminate the importance of these atom types to these two shells.
Thus, $M_{1,0}$ and $M_{2,0}$ are the atom count in these two shells
and both could serve as basis functions.
$M_{i,1}$ are vectorial quantities indicating eccentricity of these
shells: if $M_{i,1}=0$ then the $i$th shell is symmetric (to the first
order) while $M_{i,1}\ne 0$ indicate that there are ``more atoms'' in
the direction $M_{i,1}$ than in the opposite direction.

As vectorial quantities, $M_{i,1}$ are not valid basis functions, however, the valid ones are $M_{i,1}\cdot M_{i,1}$ indicating the magnitude of eccentricity and $M_{i,1}\cdot M_{i,2}$ indicating how these two eccentricities are aligned with respect to each other.
One can make many more basis functions from these quantities, e.g., $M_{i,0} (M_{i,1}\cdot M_{i,1})$, $(M_{i,1}\cdot M_{i,1}) (M_{i,1}\cdot M_{i,2})$, etc.
One can then continue by analogy: $M_{i,2}$ are the second moments of inertia of these shells indicating the degree to which these shells are ``squeezed'' in the respective directions, forming the basis functions $M_{i,2}\!:\!M_{j,2}$, $(M_{i,2} M_{j,1})\cdot M_{k,1}$, $(M_{i,2} M_{j,2} M_{k,1})\cdot M_{\ell,1}$, etc.
We remark that this way of enforcing symmetries in the potential is related to the ideas from Refs.~\onlinecite{kondor2018rotation-invariance,hirn2017wavelet}.

For the purpose of choosing which (out of the infinite number of) basis functions to include in the interatomic potential we define the degree-like measure, \emph{level}, of $M_{\mu,\nu}$ by ${\rm lev} M_{\mu,\nu} = 2\mu+\nu$ and the level of $B_\alpha$ obtained by contracting $M_{\mu_1,\nu_1}$, $M_{\mu_2,\nu_2}, \ldots$, as ${\rm lev} B_\alpha = (2\mu_1+\nu_1) + (2\mu_2+\nu_2) + \ldots$.
Thus, to define an MTP we choose some ${\rm lev}_{\rm max}$ and include in \eqref{eq:lin_expansion} each $B_\alpha$ with ${\rm lev} B_\alpha \leq {\rm lev}_{\rm max}$. 
Thus, by increasing ${\rm lev}_{\rm max}$ we increase the number of parameters in the potential, including the contributions of three-body, four-body, etc., terms.
In this sense, $V(\neiN)$ has a systematically improvable functional
form.

The radial functions $f_{\mu}(|\bm r_{ij}|,z_i,z_j)$ (cf.\ \eqref{eq:moments}) have the form:
\begin{align}\label{eq:rad}
f_{\mu}(\rho, z_i, z_j) &\phantom{:}= \sum_k c^{(k)}_{\mu, z_i, z_j} Q^{(k)}(\rho),
\qquad\text{where}\\\notag
Q^{(k)}(\rho) &:= T_k(\rho)(R_{\cut} - \rho)^2.
\end{align}
Here $T_k(\rho)$ are the Chebyshev polynomials on the interval $[R_{\min},R_{\cut}]$.
The term $(R_{\cut} - \rho)^2$ was introduced to ensure smoothness with respect to the atoms leaving and entering the cut-off sphere. Taking into account that in real systems atoms never stay too close to each other, we can always choose some reasonable value for $R_{\min}$. 

The difference from the single-component MTPs \cite{shapeev2016moment} is that now the functions $f_{\mu}(\rho, z_i, z_j)$ depend on the types of the central and the neighboring atoms.
As follows from \eqref{eq:rad} a number of parameters $c^{(k)}_{\mu, z_i, z_j}$ exist for each pair of species and each $\mu$. Note that the number of these parameters is proportional to $n^2$, where $n$ is the number of species, while number of parameters $\xi_i$ from \eqref{eq:lin_expansion} does not depend on the number of species. Thus, the total number of model parameters $\bm{\theta}=(\{\xi_i\},\{c^{(k)}_{\mu, z_i, z_j}\})$ to be found in the minimization procedure \eqref{eq:loss} grows less than quadratically with the number of species, despite accounting for many-body interactions in $V(\neiN)$.
It was proven \cite{shapeev2016moment} that the descriptors of the form \eqref{eq:moments} provide a complete description of an atomic neighborhood, in the sense that any function of atomic neighborhood with the same symmetries as $V(\neiN)$ can be approximated as a polynomial of these descriptors with an arbitrary accuracy. The proof \cite{shapeev2016moment} holds for a single-component case, but can be easily extended to a multicomponent case.

\begin{figure}[htbp] 
	\includegraphics[width=.35\textwidth]{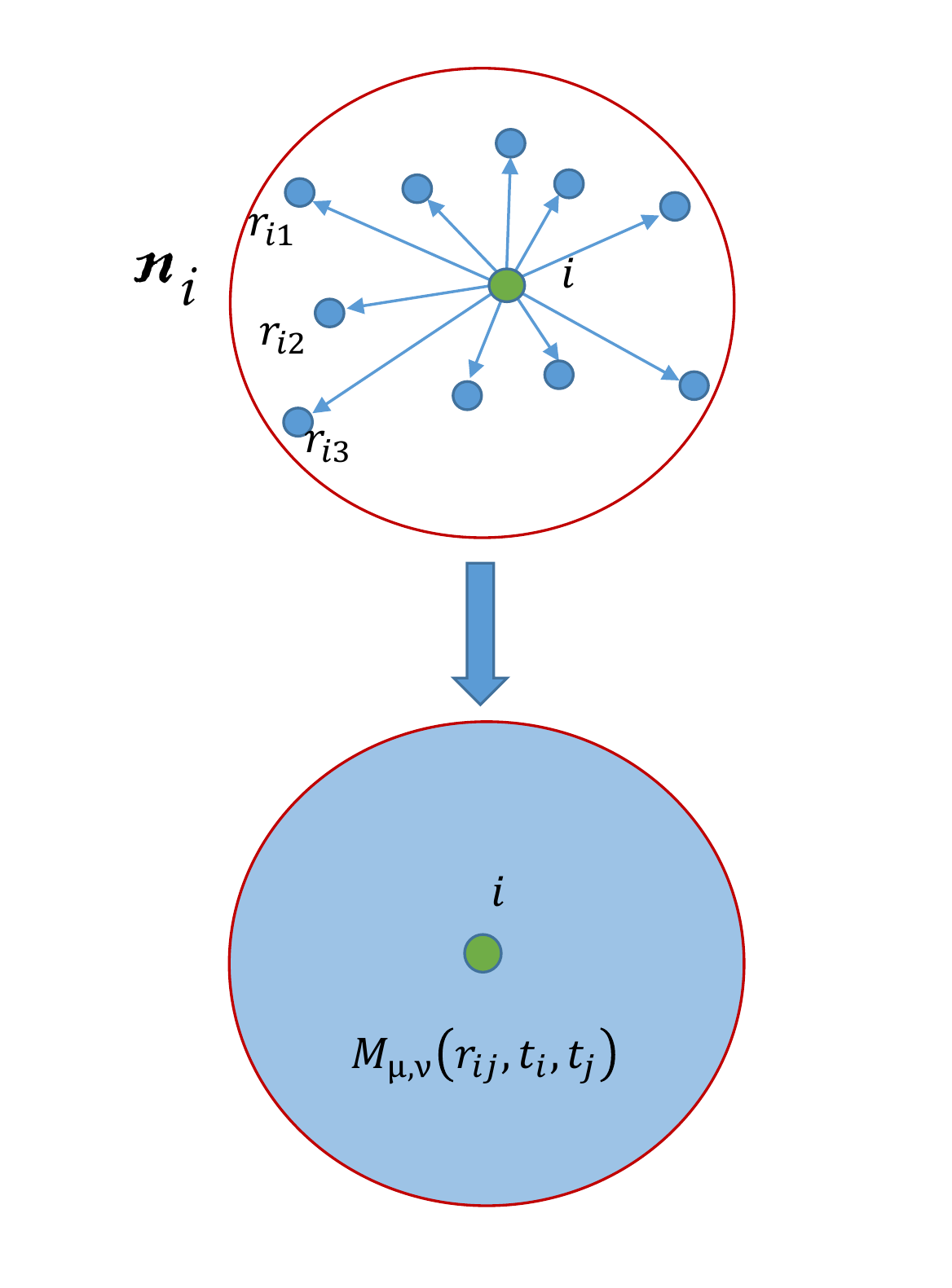}
	\caption {For the purpose of fitting the interatomic interaction energy $E$, the neighborhood $\neiN_i$ is described by the moment tensors $M_{\mu,\nu}$ exhibiting all the physical symmetries that $E$ has.}
	\label{fig:moments}
\end{figure}

\subsection {Active Learning}\label{sec:maxV}

The accuracy of a machine-learning potential depends as much on a good
	functional form (an efficient representation, in ML parlance), as on
	the quality of the training set. Roughly speaking, a good training set
	should include all the representative structures, so that the
	potential does not have to ``extrapolate'' while searching for the
	stable phases.  In cluster expansion-like approaches in which the
	energies of relaxed structures are predicted based on a
	representation that uses unrelaxed structure geometries,
	extrapolation results in higher prediction errors \cite{PhysRevB.96.014107}.
	However, in our approach structural relaxation is treated
		explicitly, and we accelerate
		the relaxation by using a machine-learning potential instead of
		DFT. Because of the added flexibility of the MTP (as compared to
		CE), avoiding extrapolation
		is even more important---it is crucial to the reliability of the
		algorithm---as highly unphysical structures can arise during the
		relaxation if the extropolation is severe.

We avoid extrapolation by using active learning.
The active learning algorithm developed in this
work is a generalization of an algorithm proposed for linearly
parametrized models \cite{podryabinkin2017active}. It is based on a
D-optimality criterion as we explain below. As the model in
this paper has a nonlinear dependence on its parameters, we
apply a
generalization of the D-optimality criterion to the nonlinear case.

To that end, we interpret fitting as solving the following overdetermined system of equations with respect to $\btheta$:
\[
E\big(\btheta,x^{(i)}\big) = E^\qm\big(x^{(i)}\big).
\]
We assume that we have some initial guess, $\bar{\btheta}$, for the optimal values of parameters.
Then if we linearize these equations around $\bar{\btheta}$ then the left-hand side will be the following tall (Jacobi) matrix
\[
\mB = \begin{pmatrix}
\frac{\partial E}{\partial \theta_1}\big(\bar{\btheta},x^{(1)}\big) & \ldots & \frac{\partial E}{\partial \theta_m}\big(\bar{\btheta},x^{(1)}\big) \\
\vdots & \ddots & \vdots \\
\frac{\partial E}{\partial \theta_1}\big(\bar{\btheta},x^{(n)}\big) & \ldots & \frac{\partial E}{\partial \theta_m}\big(\bar{\btheta},x^{(n)}\big) \\
\end{pmatrix},
\]
where each row corresponds to a particular structure from the training set.

The generalized D-optimality criterion hence states that the best training set of $m$ configurations corresponds to a square $m\times m$ submatrix $\mA$ of the matrix $\mB$ of maximal volume (i.e., with maximal value of $|{\rm det}(\mA)|$).
In practice, it is sufficient, for a given configuration $x^*$ to compute its \textit{extrapolation grade} $\gamma(x^*)$ defined as the maximal factor by which $|{\rm det}(\mA)|$ can grow if $x^*$ is added to the training set.
We do it by using the so-called maxvol algorithm \cite{goreinov2010find}, according to which
\[
\gamma(x^*) = \max_{1\leq j \leq n} (|c_j|),
\qquad\text{where}
\]
\[
c = \bigg(
\frac{\partial F}{\partial \theta_1}\big(\bar{\btheta},x^*\big) \ldots \frac{\partial F}{\partial \theta_n}\big(\bar{\btheta},x^*\big)
\bigg) \mA^{-1} =: b^* \mA^{-1}.
\]
Thus, we add $x^*$ to the training set if $\gamma(x^*) \geq \gamma_{\rm tsh}$, where $\gamma_{\rm tsh} \geq 1$ is a tunable threshold parameter that can control how much extrapolation is allowed.

The (generalized) D-optimality criterion serves to detect structures on which the potential extrapolates.
Hence, training on such structures prevents extrapolation and thus ensures that all the structures occurring during relaxation are interpolative with respect to the structures in the training set.
In this work we use the active learning algorithm to construct the training set by selecting some of the configurations arising during relaxation.
It also can be used to compose an optimal training set from configurations belonging to some pre-defined set \cite{gubaev2018machine}. 

\subsection {Algorithm} \label{sec:lotf}

\begin{figure*}[htbp] 
	\includegraphics[width=.75\textwidth]{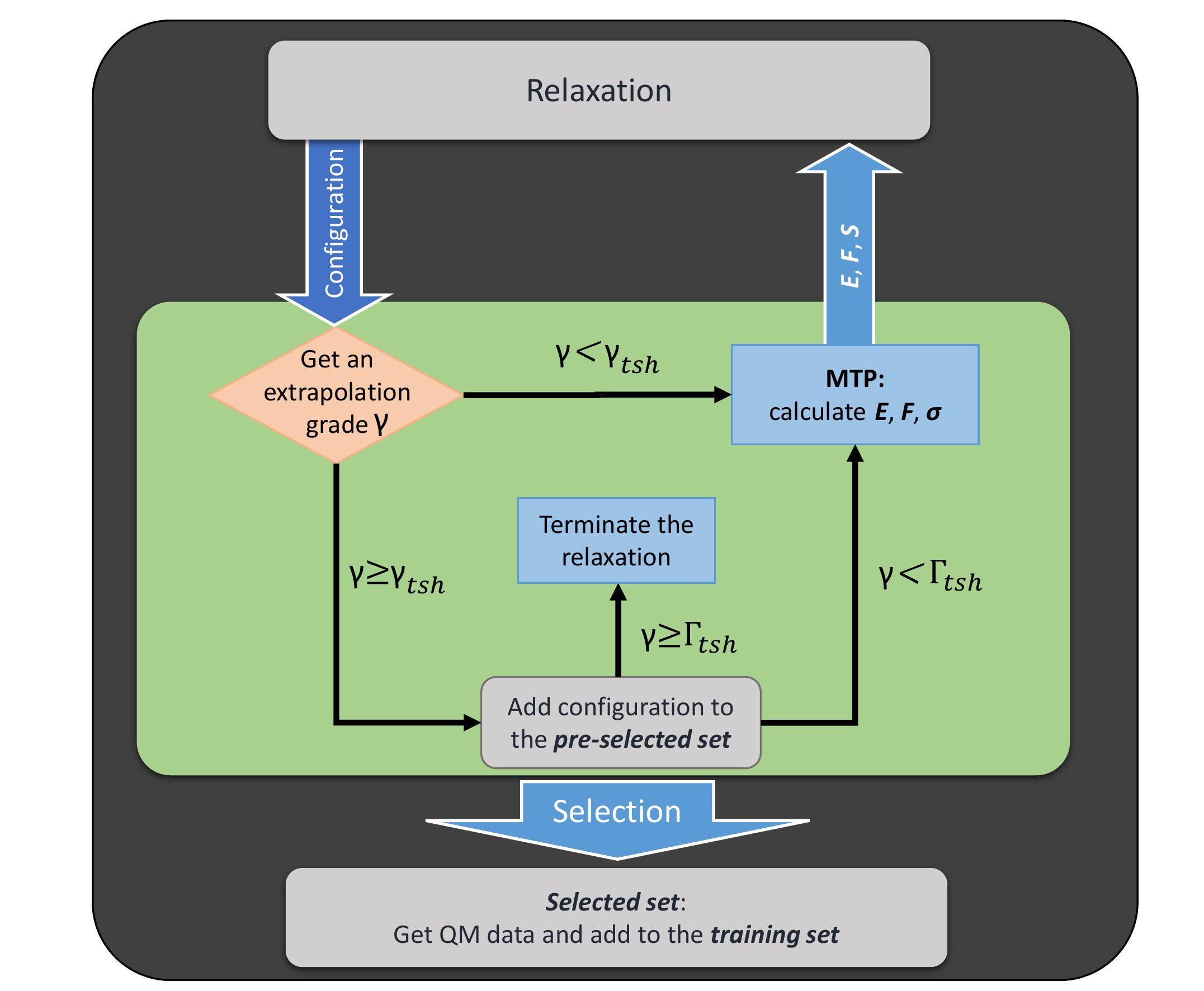}
	\caption{Relaxation with active learning. If MTP encounters an
	extrapolative configuration ($\gamma \geq \gamma_{\rm
		tsh}$), the configuration is added to the preselected set
	for further selection. In the case of significant
	extrapolation ($\gamma \geq \Gamma_{\rm tsh}$) the
	relaxation is terminated. For configurations with $\gamma <
	\Gamma_{\rm tsh}$, the MTP provides energies, forces and
	stresses. If no configuration with $\gamma \geq \Gamma_{\rm tsh}$
	is encountered, the relaxation stops at some equilibrium
	configuration.}
\label{fig:relax_scheme}
\end{figure*}

Next we describe the algorithm for constructing the convex hull.
\begin{description}
\item[Input] The input to the algorithm is:
\begin{enumerate} 
	\item A set of candidate structures among which we expect to find the groundstate structures.
	
	 (We can afford to select a much broader and more diverse set of structures as compared to the approaches based solely on DFT.)
	
	\item A functional form of MTP, $E=E(\btheta,x)$.
	
	We initialize $\btheta$ randomly and let the training set be empty.
	
	\item A quantum-mechanical model $E^\qm(x)$.
	
	In this work we used DFT as implemented in VASP 5.4.1.
	
	\item Two thresholds $\gamma_{\rm tsh}$ and $\Gamma_{\rm tsh}$, such that $\Gamma_{\rm tsh} > \gamma_{\rm tsh} > 1$.
	
	If the extrapolation grade $\gamma(x^*)$ is greater than $1$, the algorithm makes two decisions: to add $x^*$ to the training set if $\gamma(x^*) > \gamma_{\rm tsh}$ and to terminate the relaxation if $\gamma(x^*) > \Gamma_{\rm tsh}$ (assuming in the latter case that we cannot make reliable predictions of energy, forces, and stresses for $x^*$), as explained below.
	
\end{enumerate}
	
\item[Step 1] For each candidate structure we perform the structure relaxation with the current MTP (defined by the current values of $\btheta$).
There can be two outcomes of the relaxation: (1) the relaxation completed successfully and we get an equilibrium structure as a result, (2) the relaxation was not successful because we encountered a structure on which the MTP attempted to extrapolate.
More precisely, the following scenarios can emerge:
\begin{itemize}
	\item[a.] The relaxation successfully converges to an equilibrium configuration and on each configuration from the relaxation trajectory the MTP does not significantly extrapolate, i.e., the extrapolation grade of each intermediate configuration is less than $\Gamma_{\rm tsh}$.
	During the relaxation there could be, however, configuration with extrapolation grade exceeding $\gamma_{\rm tsh}$---in this case we add such a configuration to the \emph{preselected set} (see Figure \ref{fig:relax_scheme} and Section \ref{sec:meth}).
	
	\item[b.] At some step of the relaxation we obtain a configuration with the extrapolation grade exceeding $\Gamma_{\rm tsh}$. This means that MTP cannot provide a reasonable prediction as it extrapolates significantly on this configuration and needs to be retrained with more ab-initio data. We then terminate the relaxation. The last and all the previous configurations with the grade exceeding $\gamma_{\rm tsh}$ are added to the \textit{preselected set}. 	
\end{itemize}

\item[Step 2] Out of the preselected set from the step 1b, we select a smaller number of configurations that will be added to the training set.
The preselected set can be very large and contain hundreds of thousands configurations (note that during the first iteration of the algorithm all the relaxations will be terminated according to the scenario (b), as the training set is empty and the MTP extrapolates on every configuration.)
Therefore we use our active learning algorithm to select up to few hundred most representative configurations, according to the D-optimality criterion from Section \ref{sec:maxV}. Thus, we extend the training domain of the MTP as much as possible while keeping the amount of \textit{ab initio} calculations relatively small. 
After the calculation of \textit{ab initio} energies, forces and stresses the selected configurations are added to the training set.

\item[Step 3] Fit the MTP on the updated training set. 
As the size of the training set grows on each iteration of the algorithm, this step will take more and more time during each subsequent iteration, but still this time is a small fraction of the time spent on {\it ab initio} calculations.

\item[Step 4] Repeat the steps 1--3, unless all the relaxations have successfully converged to the respective equilibrium configurations.
\end{description}
As we keep refitting the MTP during the relaxation on a dynamically updated training set, we call this algorithm as ``relaxation while learning on-the-fly''. 

\begin{figure}[htbp]
	\subfigure[]{\includegraphics[width=\linewidth]{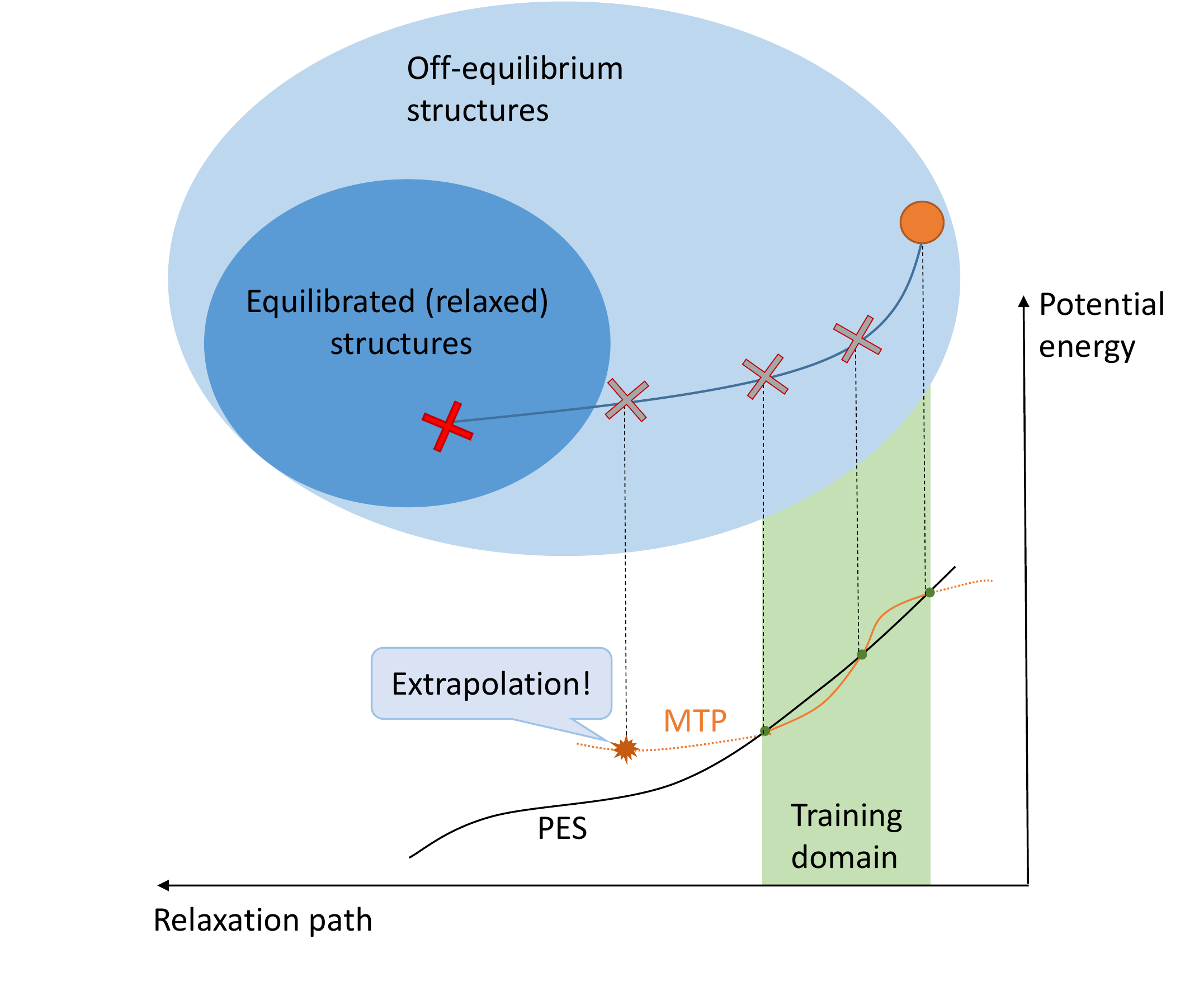}}
	\subfigure[]{\includegraphics[width=\linewidth]{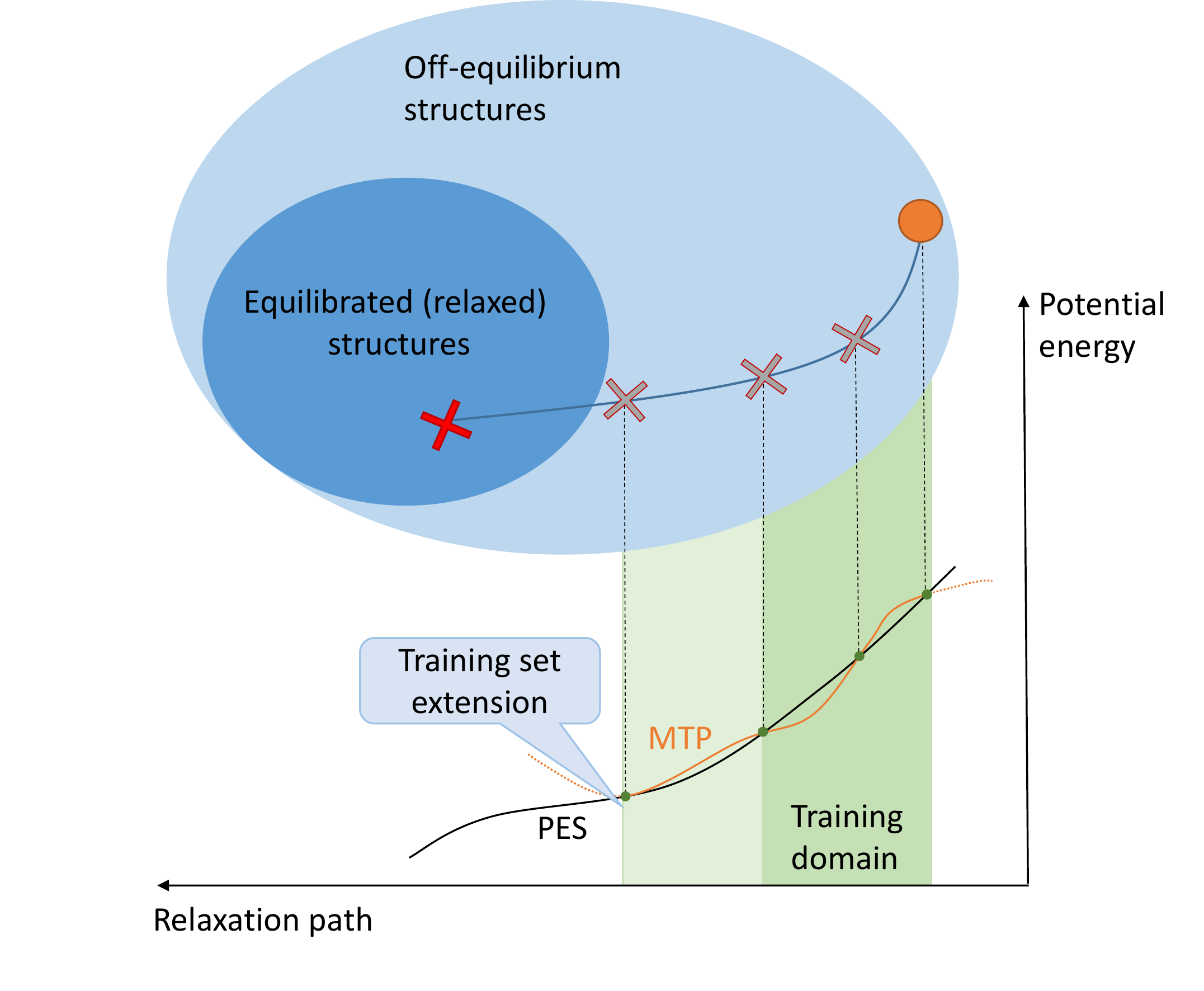}}
	\caption {If an MTP encounters some extrapolative configuration during relaxation, as shown in (a), the relaxation is terminated and restarted after retraining the MTP on this configuration, as shown in (b).}
	\label{fig:ts_extension}
\end{figure}

\begin{figure}[htbp] 
	\includegraphics[width=\linewidth]{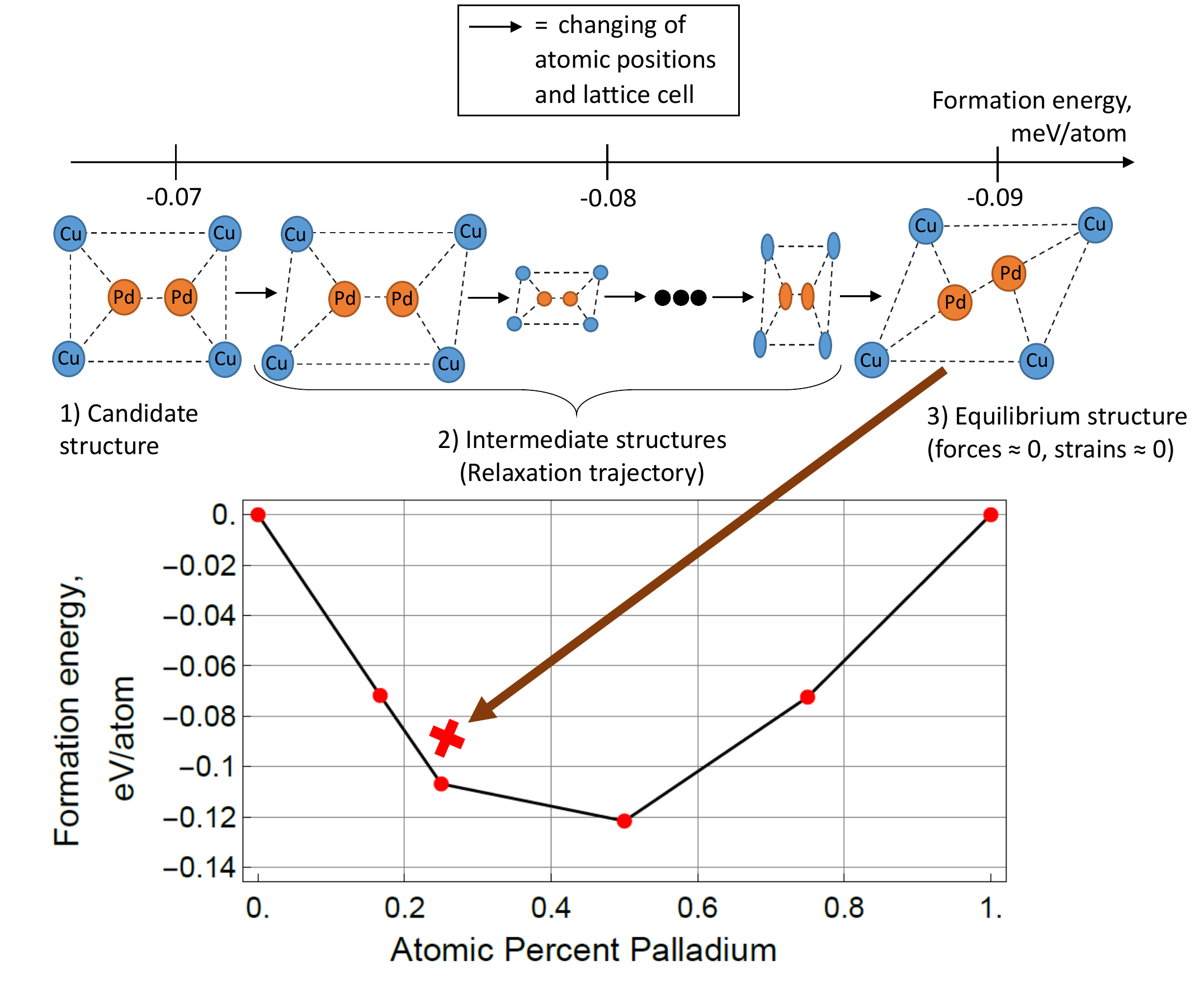}
	\caption {Graphical illustration of the relaxation process. By
          ``relaxation trajectory'' we mean a sequential list of
          structures that occur during the relaxation which have
          similar but distinct atomic displacements and lattice
          parameters, and which typically have decreasing energy.}
	\label{fig:relax_traj}
\end{figure}

\section{Results and Discussion}\label{sec:results-and-discussion}

\subsection* {CuPd system} \label{sec:CuPd}

To test the applicability of our algorithm (Section \ref{sec:lotf}) to the prediction of stable alloy structures we first used it to construct the Cu-Pd convex hull.  We chose the Cu-Pd system because the structure of both pure Cu and Pd is fcc, while the stable equimolar CuPd structure is a bcc derivative structure. This system is a good test of whether or not our   MTP-based model is able to simultanously handle multiple lattice   types, a challenging case for cluster expansion.

Using the algorithm \cite{hart2012generating}, we prepared 40,000 unrelaxed configurations with the bcc lattice and close-packed latticess (fcc and hcp), each configuration with 12 or less atoms in the unit cell and different concentrations of Cu and Pd. These were the candidate structures served as the input to our relaxation while learning-on-the-fly algorithm.  We then equilibrated them and constructed a convex hull based on their relaxed energies.  As follows from the scheme from Section \ref{sec:lotf}, the training set increases on each iteration. The final training set was formed by 523 configurations. We call the training set ``final'' since an MTP trained on this set is able to relax all the candidate structures without exceeding threshold for the extrapolation grade.  The RMSE of energy per atom ($\sigma$) measured on this training set was equal to $2.3$ meV/atom.  We used ${\rm lev}_{\rm max} = 16$ (refer to Section \ref{sec:meth:mlip}) to construct the MTP with about 200 parameters $\btheta$.

\begin{figure*}[htbp]
	\hfill
	\subfigure[]{\includegraphics[width=.48\textwidth]{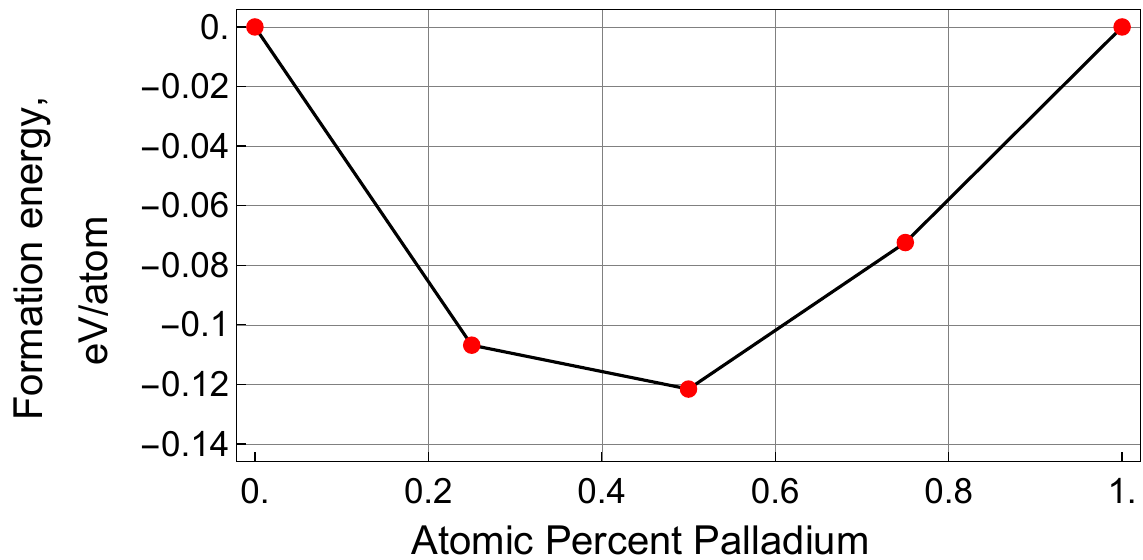}}
	\hfill
	\subfigure[]{\includegraphics[width=.48\textwidth]{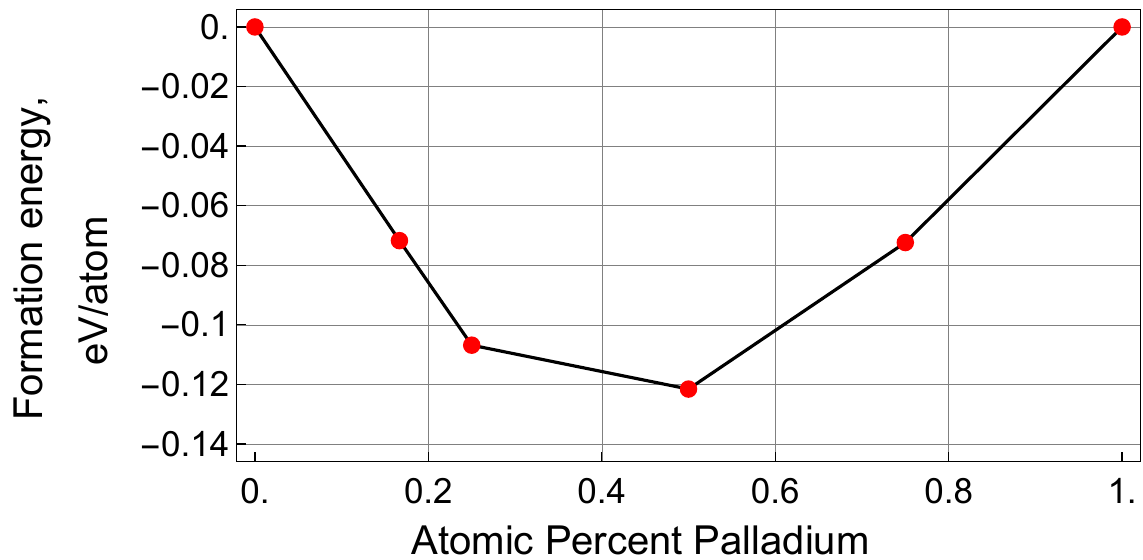}}
	\hfill$\mathstrut$
	\caption {Comparison of the convex hulls (a) as obtained from AFLOW and re-calculated with DFT, and (b) as found by MTP and re-calculated with DFT. We have discovered a structure at 16.6\% Pd which is 0.5 meV lower than AFLOW's convex hull.}
	 \label{fig:ch_compare}
\end{figure*}

Figure \ref{fig:ch_compare} shows the convex hulls constructed by the MTP and by high-throughput DFT calculations as reported in AFLOW.
To make a direct comparison possible, both convex hulls were post-relaxed with DFT using the same settings (such as pseudopotentials, k-point mesh, etc.).
As a result, we have found a structure with 16.6\% concentration of Pd that is not presented in the AFLOW library and has energy per atom 0.5 meV below the AFLOW convex hull level.
Though such a shallow ground state is typically not significant beyond academic interest,   Cu-rich ground states are believed to have an  effect on the   experimental Cu-Pd phase   diagram and have been discussed in Refs.~\onlinecite{barthlein2007reinterpreting,barthlein2009stability} as a way of explaining the     peculiar ``off-stoichiometry'' behavior on the Cu-rich side of the phase diagram.

\begin{figure}[htbp] 
	\includegraphics[width=.45\textwidth]{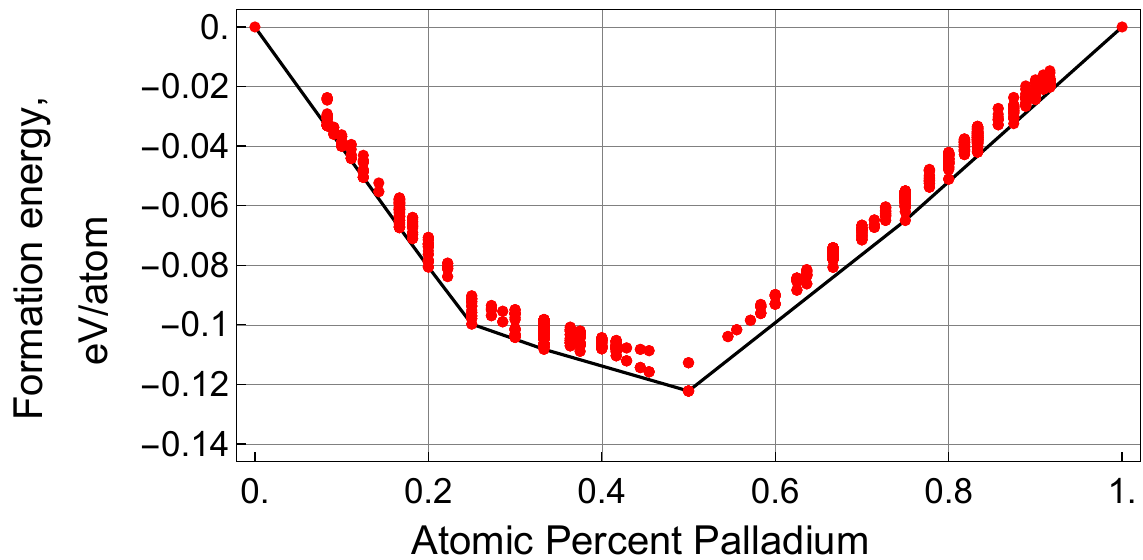}
	\caption {Convex hull constructed by MTP and structures with formation energy within 10 meV/atom above the convex hull.}
	\label{fig:ch_interv}
\end{figure}

It is illustrative to show the convex hull predicted by MTP and not post-relaxed with DFT. In Figure \ref{fig:ch_interv}, only structures within the 4$\sigma$ (10 meV/atom) interval from the MTP convex hull are shown.
Visually, the MTP convex hull looks slightly different due to the approximation errors of MTP different relative levels of the structures on the ``energy per atom'' axis.
Still, MTP reproduced the stable phases present in AFLOW library.

During the entire procedure, most of the computational expense (about 90\%) was DFT calculations. In total, we did 523 single-point DFT calculations.  If we relaxed all the 40,000 configurations using DFT, it would have taken about 10,000 times more computing time.

\subsection* {Co-Nb-V}

We next test our algorithm on constructing a convex hull for the ternary Co-Nb-V system in the region where the concentration of Co is 50\% or more. 
The number of initial candidates was about 27,000 and they were bcc-like and close-packed (fcc, hcp, etc.) configurations with 8 or less atoms in the unit cell and different concentrations of Co, Nb and V.

\begin{figure}[htbp]
	\includegraphics[width=.50\textwidth]{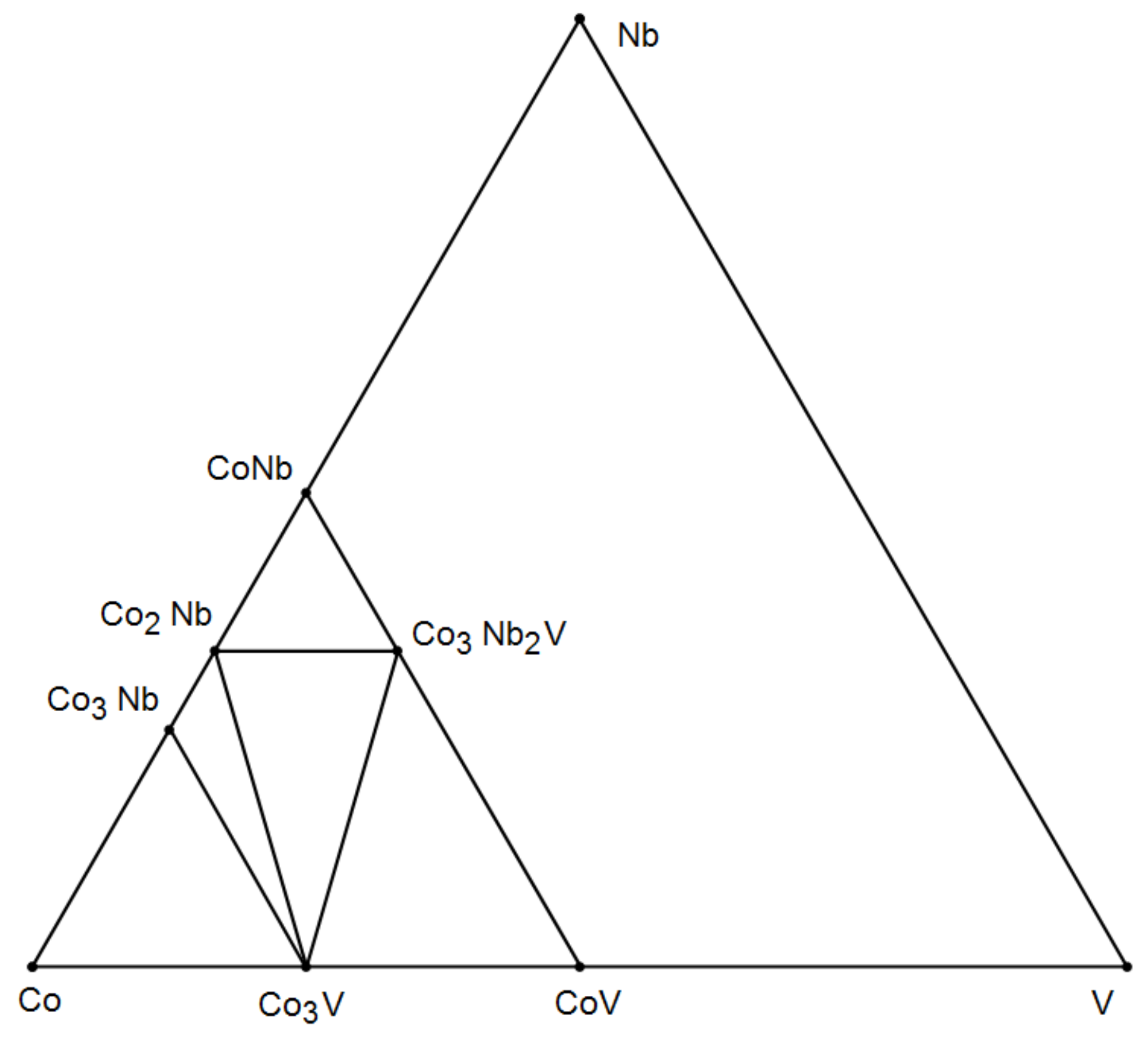}
	\caption {Convex hull of the Co-Nb-V system constructed by MTP in the Co-rich region.}
	 \label{ch_tern}
\end{figure}

\begin{figure*}[htbp]
	\subfigure[]{\includegraphics[width=.45\textwidth]{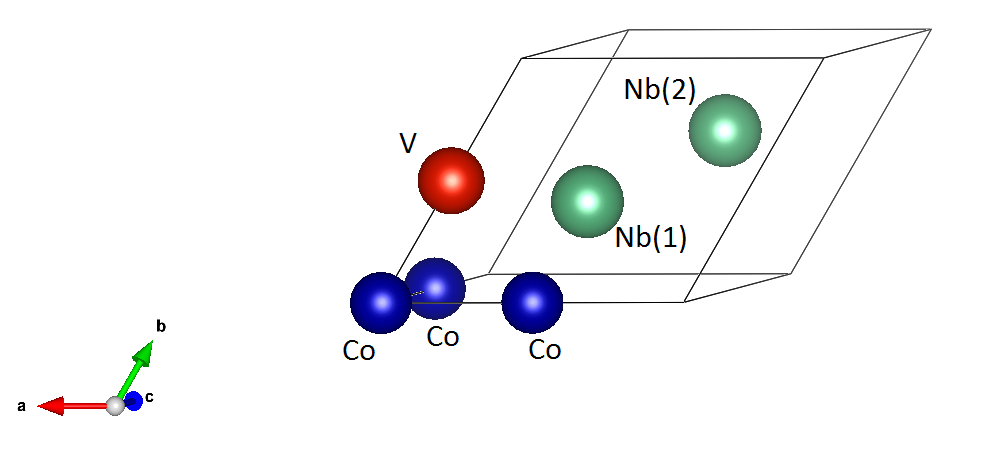}}
	\\
	\subfigure[]{\includegraphics[width=.45\textwidth]{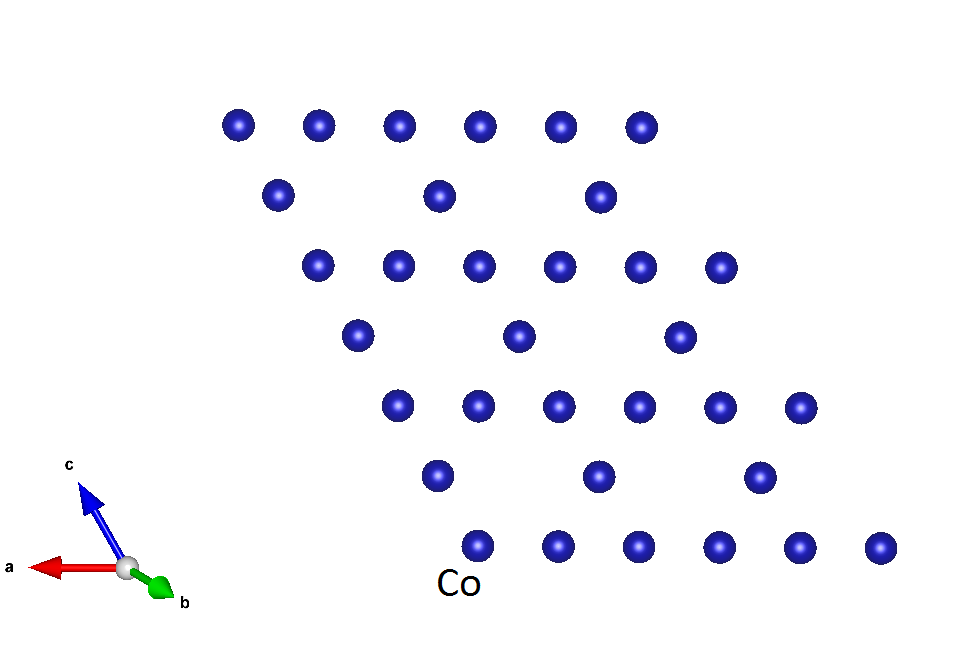}}
	\subfigure[]{\includegraphics[width=.45\textwidth]{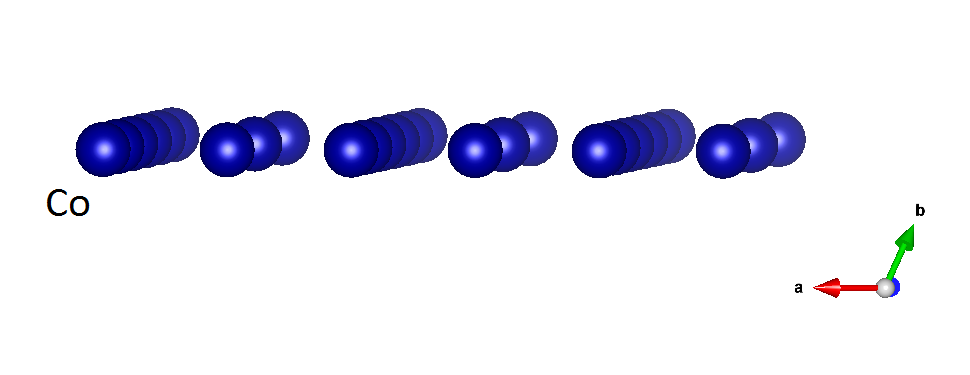}}
	\subfigure[]{\includegraphics[width=.45\textwidth]{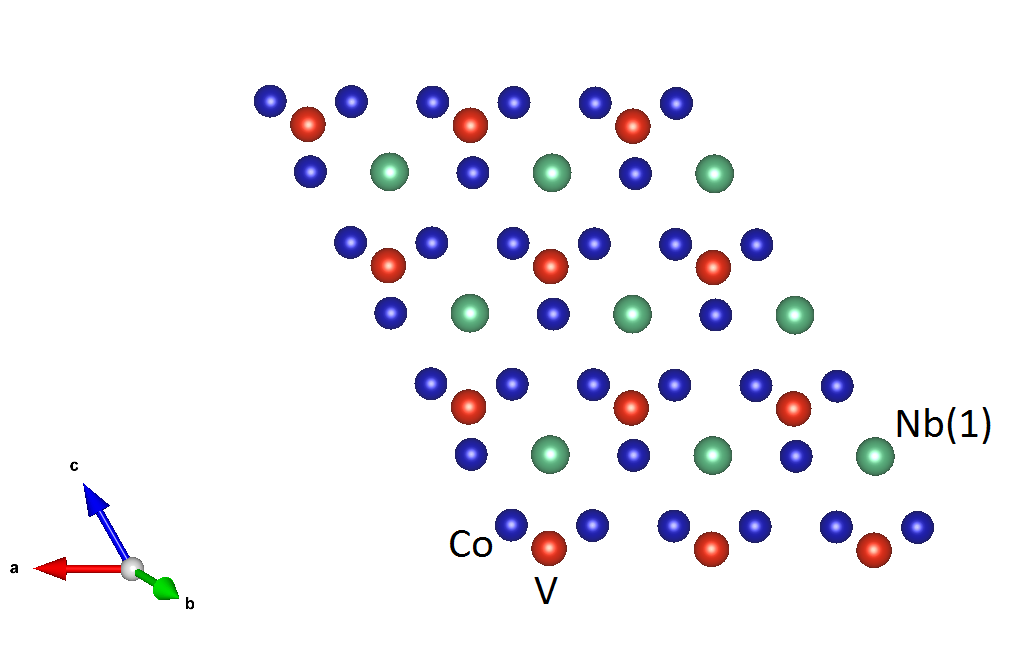}}
	\subfigure[]{\includegraphics[width=.45\textwidth]{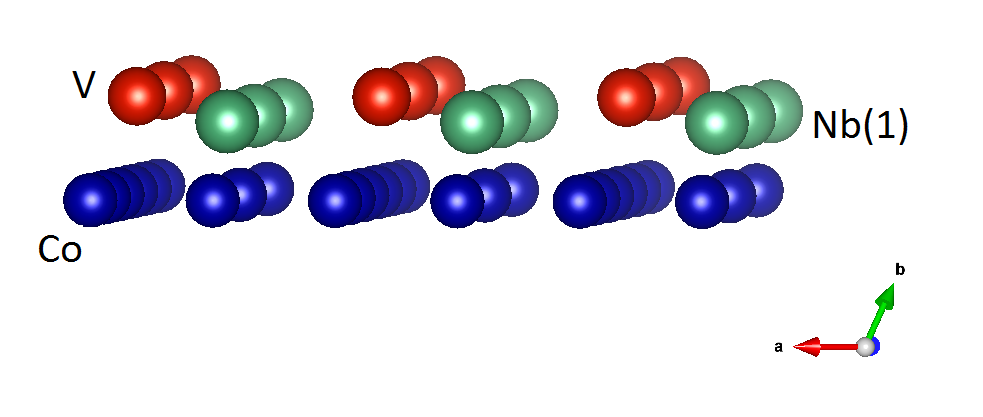}}
	\subfigure[]{\includegraphics[width=.45\textwidth]{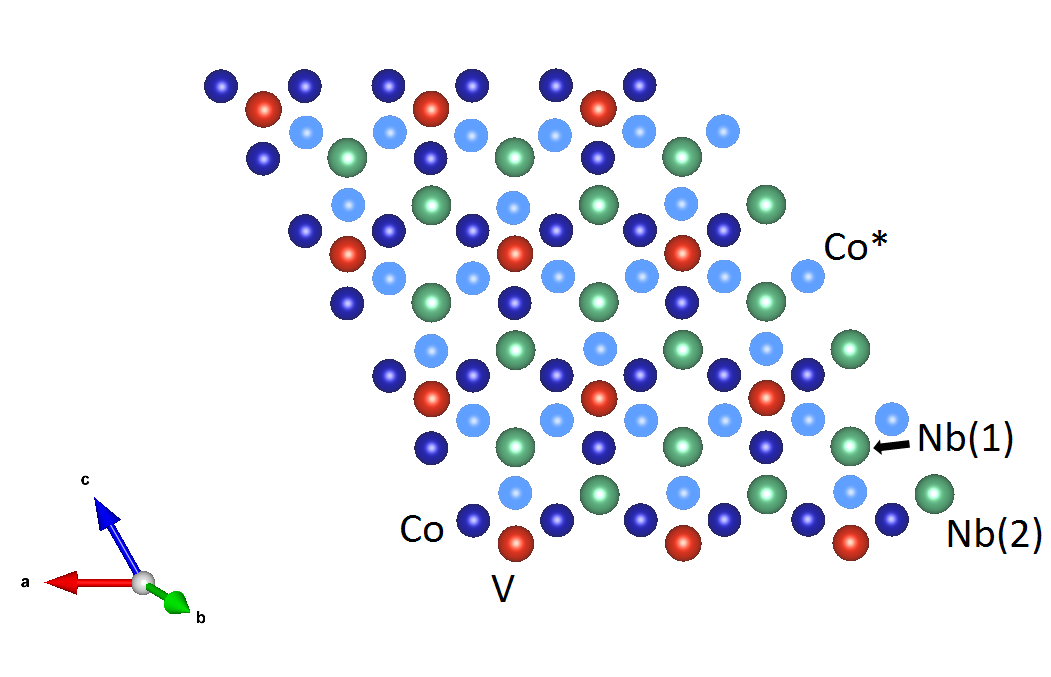}}
	\subfigure[]{\includegraphics[width=.45\textwidth]{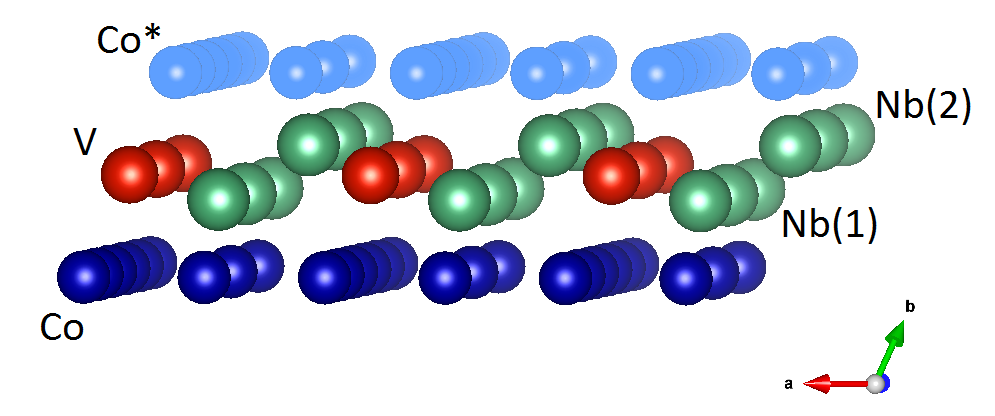}}
	\caption{The Co$_3$Nb$_2$V discovered by MTP.
			The unit cell is shown in (a), while layer-by-layer plots in vertical and side projections are shown in (b)--(g).
			Co* show were the next (periodically extended) layer of Co atoms are positioned.
			The structure was found, although no similar crystal prototypes were used.
	}
	\label{fig:conbv_layers}
\end{figure*}

The MTP was trained on-the-fly and the final training set consisted of
383 configurations.  The resulting convex hull is shown in Figure
\ref{ch_tern}.  Remarkably, we have discovered a new structure with
composition Co$_3$Nb$_2$V.  It has the formation energy of $50$
meV/atom below the AFLOW convex hull level.  Its unit cell and a
layer-by-layer plot are shown in Figure \ref{fig:conbv_layers}.
We remark that geometrically this structure is different from any
  of those in the initial pool---e.g., the Nb atoms have 16 nearest
  neighbors with distances between 2.76 and 2.98 \AA.  It would hence
  be impossible to accurately treat such a configuration for both an
  on-lattice model, such as cluster expansion, and an off-lattice
  model unless such a crystal prototype was known and explicitly added
  to the training set.  This demonstrates the
  capabilities of our approach, combining an accurate off-lattice
  model and active learning.
	
\subsection* {Al-Ni-Ti} \label{sec:AlNiTi}

Finally, we applied our algorithm to the Al-Ni-Ti system.  This system
is well-studied and has many known ternary structures, some of which
have over 20 atoms in the unit cell.  We hence chose a set of
candidate structures consisting of two parts: the first part has 1463
structures which were used in AFLOW as crystal prototypes.

The second part was generated by the algorithm from Ref.~\onlinecite{hart2012generating} and contains 375,000 binary and ternary structures enumerating all possible unit cells with different
symmetries (bcc, fcc\ and hcp) and different number of atoms; we chose
unit cells containing up to 12 atoms.

Including crystal structure prototypes adds extra difficulties: the
structures may contain short interatomic distances (if, e.g., the
original structure from which the prototype was derived had
carbon-metal bonds which are shorter than typical metal-metal
distances) and also smaller volume than that of the typical Al-Ni-Ti
structures.  Both of these features of the prototypes might result in
unphysical structures with large stresses and forces which, in turn,
lead to large MTP prediction errors.  To make the unit cells of the
candidate structures less deformed, we adjusted their volumes
enforcing the relation:
\begin{equation} \label{eq_vol}
v(n_{\mathrm{Al}},n_{\mathrm{Ni}},n_{\mathrm{Ti}}) = n_{\mathrm{Al}}v_{\mathrm{Al}} +  n_{\mathrm{Ni}}v_{\mathrm{Ni}} +  n_{\mathrm{Ti}}v_{\mathrm{Ti}},
\end{equation}
where $v(n_{\mathrm{Al}},n_{\mathrm{Ni}},n_{\mathrm{Ti}})$ is the volume per atom assigned to the unit cell with concentrations of Al, Ni, Ti equal to $n_{\mathrm{Al}}$, $n_{\mathrm{Ni}}$, $n_{\mathrm{Ti}}$ respectively and $v_{\mathrm{Al}}$, $v_{\mathrm{Ni}}$, $v_{\mathrm{Ti}}$ are the volumes per atom for equilibrium fcc-Al, fcc-Ni, hcp-Ti structures respectively. Resizing the unit cells in this way provides an initial guess for their volumes (a kind of ``Vegard's law'' for different lattice types.)

To circumvent the large prediction errors that might occur for prototype structures with bond lengths and neighborhoods atypical of alloys, we performed a two-step relaxation as explained below. We used ${\rm lev}_{\rm max} = 20$ (see Section \ref{sec:meth:mlip}) to construct the MTP with about 650 parameters.  This makes the potential more accurate, but requires more data for training, than with ${\rm lev}_{\rm max} = 16$.  First, we did the same procedure as for the Cu-Pd and Co-Nb-V systems, which provided us with the training set of 2393 configurations with \textit{ab initio} energies, forces and stresses. The MTP trained on this set has MAE and RMSE of energy per atom of 18 meV/atom and 27 meV/atom, respectively.

We hence relaxed the 377,000 configurations and constructed a convex hull.
Next, we picked all the configurations whose formation energy per atom is lower than $4\sigma$ ($\approx 100$ meV) from the convex hull level. This left us with 62,000 configurations.

Second, we repeated the procedure of relaxing the 62,000 configurations on-the-fly from scratch, starting from an empty training set.
During this process a new training set with 976 structures was formed by the active learning algorithm. The MAE and RMSE on this training set was 7 meV/atom and 9 meV/atom, respectively.
This way we constructed a convex hull based on more accurate formation energies, than would be possible after the first step.    

\begin{figure}[htbp]
	\includegraphics[width=\linewidth]{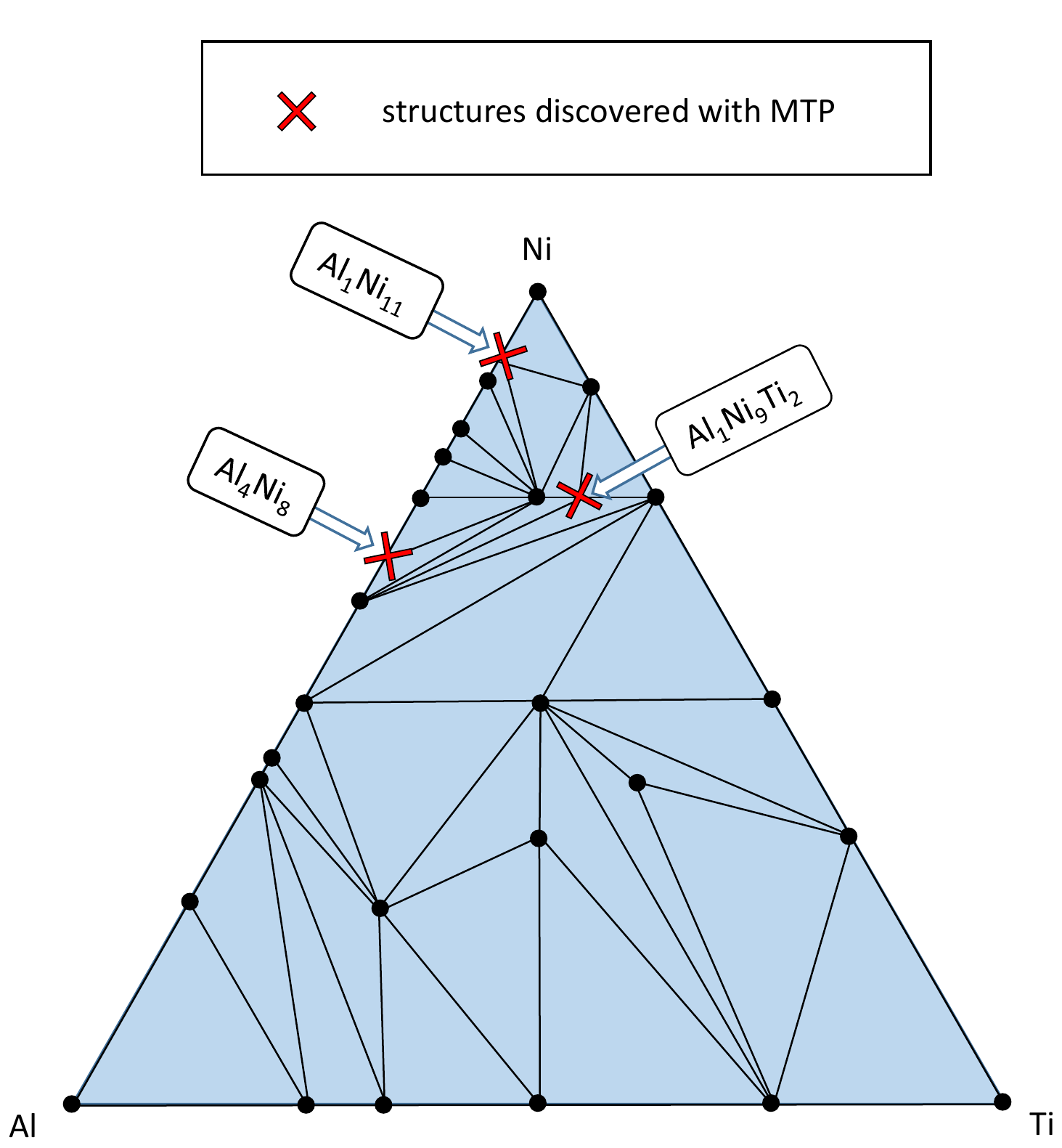}
	\caption {Al-Ni-Ti convex hull constructed by MTP and compared to the one from AFLOW.
		The MTP convex hull contains all the structures from AFLOW plus three newly discovered ones.}
	\label{fig:alniti_mtp}
\end{figure}

\begin{table}[htbp]
	\begin{tabular}{| l |c |} 
		\hline			
		Formula & Position below the convex hull \\
		\hline  
		Al$_4$Ni$_8$ &  $-7.38$ meV \\ 	
		\hline
		Al$_1$Ni$_{11}$ & $-1.18$ meV  \\ 
		\hline
		Al$_1$Ni$_9$Ti$_2$ &  $-0.34$ meV  \\ 
		\hline  
	\end{tabular}
	\caption{New Al-Ni-Ti structures found in this study. The ``level below the convex hull'' was computed using DFT.}
	\label{tab:alniti_new}
\end{table}

To perform a comparison with the AFLOW convex hull, from the 62,000 relaxed configurations we eliminated all the configurations with formation energy per atom higher than $4\sigma$ from the convex hull level, where now $\sigma=9$ meV/atom.
This left us with about 7000 configurations, which were subsequently relaxed with DFT.
After this we constructed a final convex hull using the DFT formation energies.
It has all the structures, present in AFLOW, and three new structures discovered by MTP (see Figure \ref{fig:alniti_mtp}).
Their chemical formulas are given in Table \ref{tab:alniti_new} together with their position below the AFLOW convex hull level.
Interestingly, all the structures are Ni-rich which makes their discovery relevant to the application of Ni-based alloys.

\begin{figure}[htbp]
	\includegraphics[width=\linewidth]{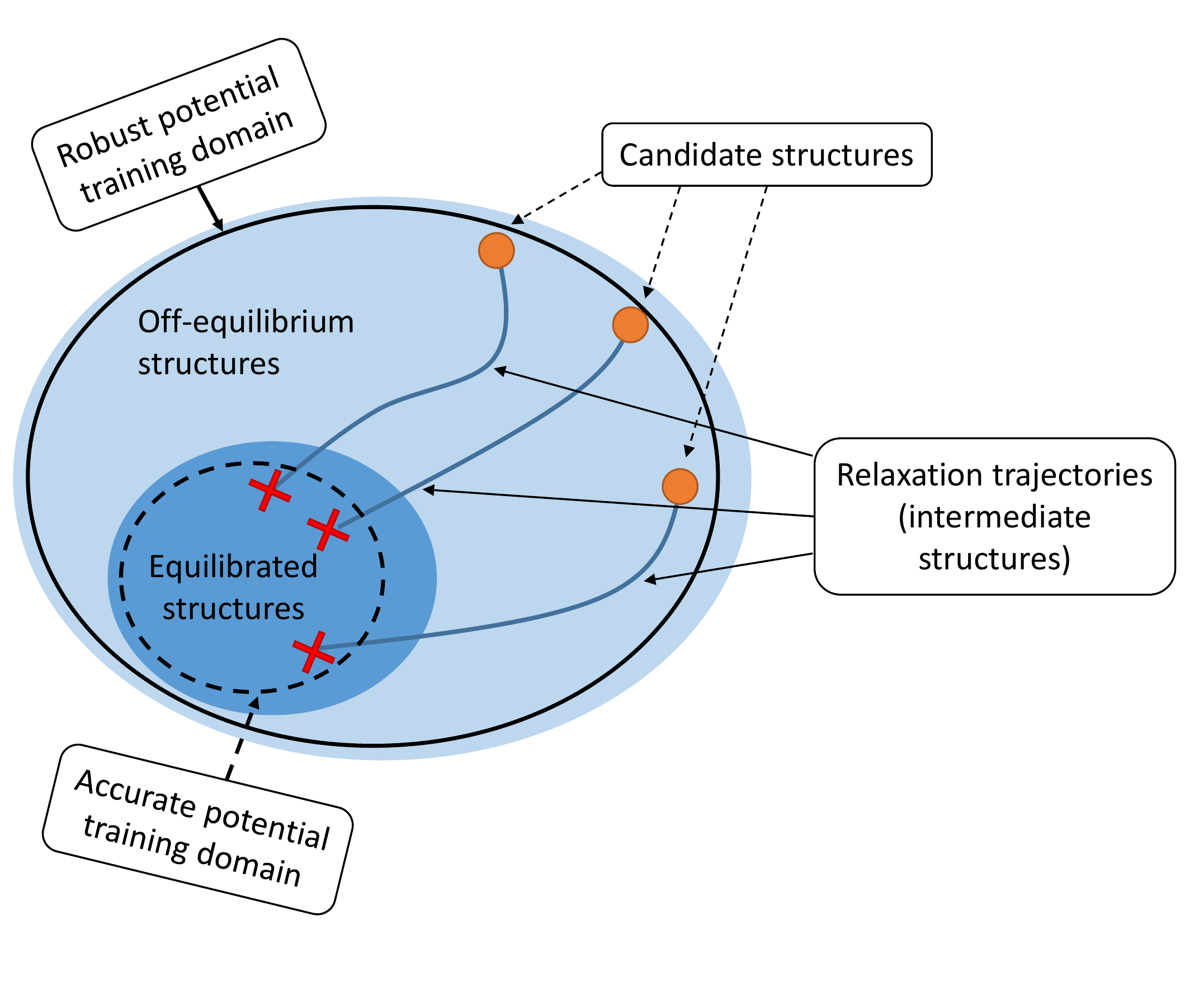}
	\caption {The accurate potential is trained on a smaller
		domain of configurational space than the
		robust one. Thus, the accurate potential provides more
		accurate predictions at the interior of the ``Relaxed structures'' region.}
	\label{fig:robust_precise}
\end{figure}

Taking into account that after the first step we have obtained an MTP capable of relaxing all the 377,000 configurations we call it the ``robust'' potential.
After the second step we have obtained an MTP which is trained on (and thus able to relax) the low-energy near-equilibrium structures only.
We refer to this MTP as the ``accurate'' potential. 
We attribute the difference in accuracies of the robust and accurate potentials to the fact that at the second step the trajectories of relaxations started from near-equilibrium structures (within the accuracy of the robust potential), see an illustration in Figure \ref{fig:robust_precise}.
This reduces the region in the configurational space in which the MTP is fitted, thus improving the accuracy in comparison to the first step.

\section{Conclusions}\label{sec:conclusions}
We have developed an algorithm for constructing a convex hull of stable alloy structures based on the moment tensor potentials (MTPs) to approximate {\it ab initio} energies, forces and stresses of atomistic configurations.
This way the calculations for atomistic systems can be done much faster than with DFT, while the accuracy is comparable to that of DFT.
The active-learning algorithm forms a training set automatically, removing the need for its manual design---the most tedious part of application of ML to atomistic modeling.
We have verified the applicability of our algorithm by constructing the convex hulls for the Cu-Pd, Co-Nb-V and Al-Ni-Ti metallic alloy systems and comparing them to the convex hulls from AFLOW library. 
For all the systems we have discovered new stable structures, which are not listed in the AFLOW library.
We attribute this to the large amount of candidate structures (40,000 for Cu-Pd, 27,000 for Co-Nb-V, 377,000 for Al-Ni-Ti) we explored, which would be impossible to equilibrate using DFT.
Instead, we performed relaxations using fast MTP calculations, referring to DFT only for the training data generation. In the cases covered by this paper the amount of single-point DFT calculations was about 1\% of the total amount of relaxed configurations.
In comparison to the high-throughput DFT calculations, the speedup is three to four orders of magnitude.

\section{ACKNOWLEDGMENTS}
The work of K.G.,\ E.V.P.,\ and A.V.S.\ was supported by the Russian Science Foundation (grant number 18-13-00479).
 G.L.W.H. was supported by the Office of Naval Research (MURI N00014-13-1-0635). This work was performed, in part, by A.V.S. at the Center for Integrated Nanotechnologies, an
Office of Science User Facility operated for the U.S. Department of Energy (DOE) Office of Science by
Los Alamos National Laboratory (Contract DE-AC52-06NA25396) and Sandia National Laboratories
(Contract DE-NA-0003525).
This collaboration has started during the MPS2016 long program at the Institute of Pure and Applied Mathematics, UCLA.

\section{Data availability}

The data required to reproduce our results are available to download from \url{http://gitlab.skoltech.ru/kgubaev/Data_for_MTP_with_active_learning}.

\bibliography{paper}

\end{document}